\documentclass[aps,prl,twocolumn,superscriptaddress,showpacs=false,floatfix]{revtex4-2}
\usepackage{amsmath,amssymb,graphicx,bm,microtype}
\usepackage[colorlinks=false,hidelinks]{hyperref}
\begin{document}
\title{Divergent Interfacial Lifetimes without Bulk Criticality in Fully Miscible Fluids}
\author{Heon Sang Lee}\email{heonlee@dau.ac.kr}
\affiliation{Department of Chemical Engineering (BK-21 Four), Dong-A University, Busan 49315, Republic of Korea}
\author{Ju Hyun Yang}\affiliation{Department of Chemical Engineering (BK-21 Four), Dong-A University, Busan 49315, Republic of Korea}
\begin{abstract}
A fully miscible interface must disappear at equilibrium, but its lifetime need not be governed by a bulk instability. We construct a spatially resolved, dissipative two-field model in which a prepared correlation-excited interface undergoes a localized saddle-node while its homogeneous excited surroundings remain metastable and the relaxed mixture is strictly one phase. The interfacial soft mode produces a deterministic first-passage time that grows as the inverse square root of the distance from the fold and couples directly to the evolving concentration-profile width. The quartic structural potential admits an exact reduction from competing positive harmonic energies on a constrained configuration manifold; sixth- and eighth-order examples provide additional numerical robustness tests. The result is an interface-specific kinetic singularity without equilibrium bulk criticality, subject to explicitly stated microscopic and continuum limits.
\end{abstract}
\maketitle
\textit{A kinetic singularity at a stable-bulk interface.---}
The homogeneous equilibrium of a fully miscible binary fluid is fixed by thermodynamics; the time required for a prepared interface to disappear is not. Transient effective interfacial tension has been reported in miscible liquids~\cite{truzz2016,truzz2017}. Interfacial persistence in reactive and compositionally evolving miscible systems provides a separate experimental motivation~\cite{pojman2006,zoltowski2007}. Microfluidic studies probe the dynamics of miscible interfaces under flow~\cite{cubaud2020}, whereas core--sheath and capillary-thinning measurements constrain their persistence and transport~\cite{hussain2024,hussain2025}. These observations do not themselves establish a singular lifetime. The sharper question is whether an interfacial lifetime can become singular without an accompanying bulk instability.

\begin{figure*}[t]
\centering\includegraphics[width=0.82\textwidth]{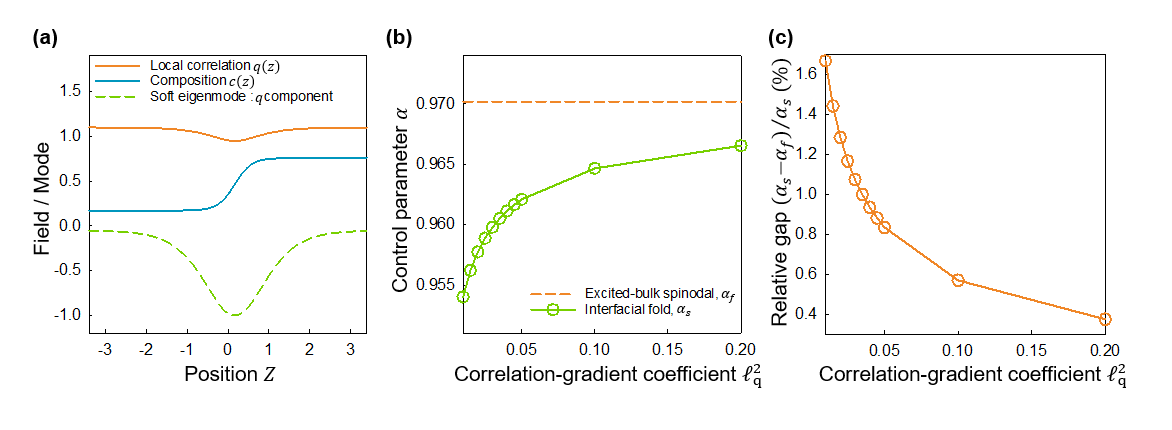}
\caption{Interface-localized loss of metastability. (a) Coupled stationary composition and correlation profiles and the localized correlation soft mode. (b) Interfacial fold parameter versus correlation-gradient stiffness, compared with the independently determined excited-bulk spinodal (dashed line). (c) Relative interface--bulk threshold gap.}\label{fig:fold}
\end{figure*}

The co-evolving variational first principle (CVFP) couples a quasi-stationary concentration profile to an evolving excess-enthalpy landscape, with a Fermi--Dirac profile obtained for its specific closure~\cite{choi2026cp}. The symmetry-field formulation of Ref.~\cite{choi2026jcp} motivates an exchange-invariant excess interaction proportional to $[c(1-c)]^2$. We retain both ingredients, but promote the excess-interaction amplitude from a collective landscape parameter to a spatially relaxing field. Unlike conventional bulk critical slowing~\cite{hohenberg1977,siggia1976,pelissetto2002,das2006}, the resulting singularity resides in a mode localized at a prepared interface. Its exponent is the familiar saddle-node exponent; its location and its coupling to a conserved concentration profile are the physical findings.

\textit{Local implementation and equilibrium constraint.---}
Let $0<c(z,t)<1$ denote conserved composition, $q(z,t)\geq0$ a nonconserved exchange-even correlation amplitude, and $p(c)=c(1-c)$. On $-H\leq z\leq H$, take
\begin{align}
\mathcal F[c,q]=\int_{-H}^{H}\!dz\bigg[&f_{\rm FH}(c;\chi)+a q p(c)^2+U(q;\alpha)\nonumber\\
&+\frac{\kappa}{2}(\partial_z c)^2+\frac{\ell_q^2}{2}(\partial_zq)^2\bigg],\label{eq:F}
\end{align}
where $f_{\rm FH}=c\ln c+(1-c)\ln(1-c)+\chi c(1-c)$ is the symmetric Flory--Huggins density~\cite{flory1942,huggins1941}, with square-gradient regularization~\cite{modica1987}, and
\begin{equation}
U(q;\alpha)=\frac{\alpha}{2}q^2-\frac23q^3+\frac14q^4.\label{eq:U}
\end{equation}
We use $\chi=1.90$, $a=1$, $\kappa=0.06$, $\ell_q^2=0.05$ unless specified otherwise. The constraint $q\geq0$ is essential: $q$ is an amplitude, not a signed order parameter. The invariant $p(c)^2$ is motivated by the exchange-symmetric contribution of the symmetry-field formulation~\cite{choi2026jcp}; the $q$ potential and its dynamics are new constitutive ingredients (Supplemental Material, Secs.~S1.1 and S20). The connection to CVFP is explicit: holding the structural field fixed during the $c$ variation, at fixed total composition, gives
\begin{align}
\frac{\delta\mathcal F}{\delta c}&=\frac{\partial h(c;q)}{\partial c}-\kappa c_{zz}=\mu,\nonumber\\
h(c;q)&=f_{\rm FH}(c;\chi)+aq[c(1-c)]^2.\label{eq:cvfp}
\end{align}
This has the CVFP quasi-stationary Euler--Lagrange form, with $q(z,t)$ providing the evolving conditional excess-enthalpy amplitude. For spatially varying $q$, this is a generalized local CVFP-type equation, not the exact original one-field closure. For a spatially uniform $q(t)$, Eq.~\eqref{eq:cvfp} has the one-field variational form with an instantaneous landscape $h(c;t)$. For spatially varying $q(z,t)$, Eq.~\eqref{eq:cvfp} remains the concentration Euler--Lagrange equation of the coupled functional; the structural Euler--Lagrange equation, Eq.~\eqref{eq:ELq}, determines stationary $q$ profiles, while the separate kinetic equation, Eq.~\eqref{eq:dyn}, governs their time evolution. The earlier exact logistic solution and its $\Lambda_c=32$ threshold apply to the original CVFP closure; neither is imposed on this distinct two-field functional (Supplemental Material, Secs.~S12--S13 and S20).

For $\alpha\geq8/9$, completing the square yields
\begin{equation}
U=q^2\left[\frac14\left(q-\frac43\right)^2+\frac\alpha2-\frac49\right]\geq0.\label{eq:positive}
\end{equation}
Because $a q p^2\geq0$, the relaxed state has $q_{\rm eq}=0$ and $f_{\rm eq}=f_{\rm FH}$. Furthermore,
\begin{equation}
f_{\rm eq}''(c)=\frac1{c(1-c)}-2\chi\geq0.20>0.\label{eq:convex}
\end{equation}
Thus the equilibrium mixture is strictly convex and one phase, even when a prepared $q$-high branch has a conditionally segregating composition landscape. This distinguishes a metastable preparation from equilibrium coexistence (Supplemental Material, Secs.~S2--S3).

\textit{A fold confined to the interface.---}
The coupled Euler--Lagrange equations, with conserved chemical potential $\mu$, are
\begin{align}
\mu&=\ln\frac{c}{1-c}+\chi(1-2c)+2aqp(1-2c)-\kappa c_{zz},\label{eq:ELc}\\
0&=ap^2+\alpha q-2q^2+q^3-\ell_q^2q_{zz}.\label{eq:ELq}
\end{align}
We solve the full equations using symmetry-pinned excited plateaux, not a prescribed logistic profile. At $H=12$ and $\Delta z=0.10$ (Supplemental Material, Secs.~S4--S5), the relaxed core force and its Schur-complement curvature vanish together at
\begin{equation}
\alpha_f=0.9620367857,\qquad q_f(0)=0.9664037684.\label{eq:fold}
\end{equation}
The independently calculated excited homogeneous-bulk spinodal is $\alpha_s=0.97019618$. At the interfacial fold the smallest homogeneous excited-bulk Hessian eigenvalue is approximately $0.20378>0$. The correlation component of the fold eigenmode has approximately $99.81\%$ of its squared weight in $|z|<2$; symmetry pinning removes the independent translation mode. Cell half-width tests at approximately fixed grid spacing and spatial-grid refinement support the separation of the local fold from bulk and finite-cell modes, without establishing an infinite-domain existence theorem (Supplemental Material, Secs.~S4--S8, including the nondegeneracy test in Sec.~S5.1).

The local mechanism follows from the positive coupling $aqp^2$: it penalizes correlation excitation most strongly in the mixed interface core. Decreasing $\ell_q^2$ from $0.05$ to $0.01$ increases $\alpha_s-\alpha_f$ from approximately $0.00811$ to $0.01618$ (Fig.~\ref{fig:fold}). Since the gradient term vanishes in a homogeneous state, this tuning acts directly on the interface without moving the excited-bulk spinodal (Supplemental Material, Secs.~S3 and S9).

\begin{figure*}[t]
\centering\includegraphics[width=0.82\textwidth]{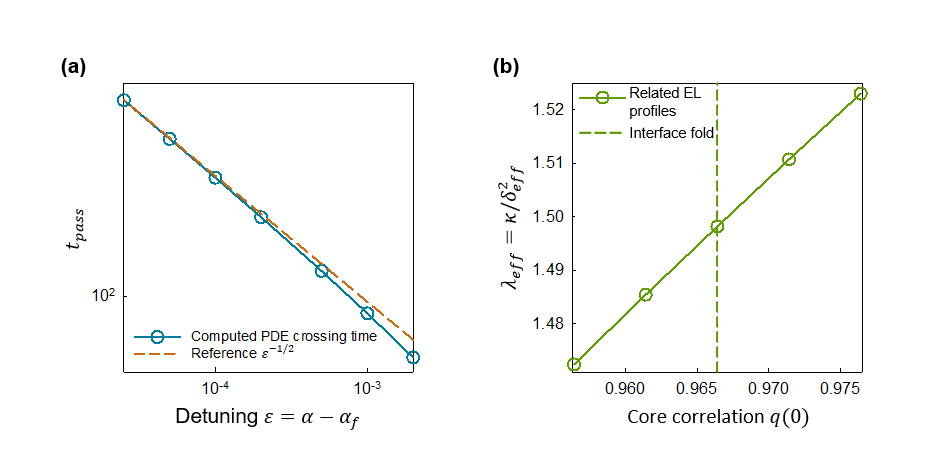}
\caption{Dynamics and CVFP width connection. (a) Full two-field PDE first-passage time to $q(0,t)=0.55$ versus positive fold detuning; dashed curve indicates the $\epsilon^{-1/2}$ reference law. (b) Operational width $\Lambda_{\rm eff}$ reconstructed from locally relaxed profiles versus constrained core amplitude.}\label{fig:dynamic}
\end{figure*}

\textit{Deterministic lifetime divergence.---}
Conserved transport and local correlation relaxation are described by
\begin{equation}
\partial_t c=\partial_z(M_c\partial_z\mu),\qquad
\partial_tq=\Pi_{q\geq0}\!\left[-M_q\frac{\delta\mathcal F}{\delta q}\right],\label{eq:dyn}
\end{equation}
with positive mobilities and an obstacle-compatible projection. In the unconstrained interior the dynamics dissipate $\mathcal F$ in the Onsager gradient-flow sense~\cite{onsager1931,doi2011} (Supplemental Material, Sec.~S10). Near a nondegenerate fold, a reduced soft coordinate $\xi$ obeys
\begin{equation}
\zeta_*\dot\xi=A(\alpha-\alpha_f)+B\xi^2+\cdots,\quad AB>0,
\end{equation}
so passage through a fixed bottleneck window on the flowing side satisfies
\begin{equation}
t_{\rm pass}\sim\frac{\pi\zeta_*}{\sqrt{AB}}(\alpha-\alpha_f)^{-1/2}.\label{eq:scaling}
\end{equation}
The exponent is not proposed as a new universality class. Full two-field simulations show persistent contrast during the tested window at $\alpha=0.961$ and core relaxation followed by mixing at $\alpha=0.965$, both below $\alpha_s$. With the core event defined by $q(0,t)=0.55$, the closest four tested detunings give an effective log--log exponent near $0.513$, approaching Eq.~\eqref{eq:scaling}; the times are dimensionless, not experimentally calibrated lifetimes (Supplemental Material, Secs.~S11--S11.1).

The singular coordinate also changes an observable profile quantity. Defining plateau contrast $\Delta c=c_+-c_-$ and operational width $\delta_{\rm eff}=\Delta c/[4|c_z(0)|]$, set $\Lambda_{\rm eff}=\kappa/\delta_{\rm eff}^2$. At the fold the numerical profiles give $\Delta c\simeq0.572$, $\delta_{\rm eff}\simeq0.2001$, and $\Lambda_{\rm eff}\simeq1.4982$. Constrained-relaxed neighboring profiles have a nonzero width response to $q(0)$ (Fig.~\ref{fig:dynamic}); they are not all independent stationary solutions. The width mapping (Supplemental Material, Secs.~S12--S13) operationalizes CVFP coevolution without imposing the original logistic solution or its model-specific rupture threshold. In the full two-field PDE tests, a second event defined by loss of endpoint composition contrast to $\Delta c=0.1$ occurs after core passage; the observed additional delay remains approximately 453--455 across the seven sampled detunings (Supplemental Material, Sec.~S11.2). This supports transmission of the bottleneck to a concentration observable without independently proving its asymptotic exponent. In particular, the concentration variation retains a CVFP-type variational structure, whereas the structural-field dynamics endogenize the evolving interfacial landscape.

\begin{figure*}[t]
\centering\includegraphics[width=0.82\textwidth]{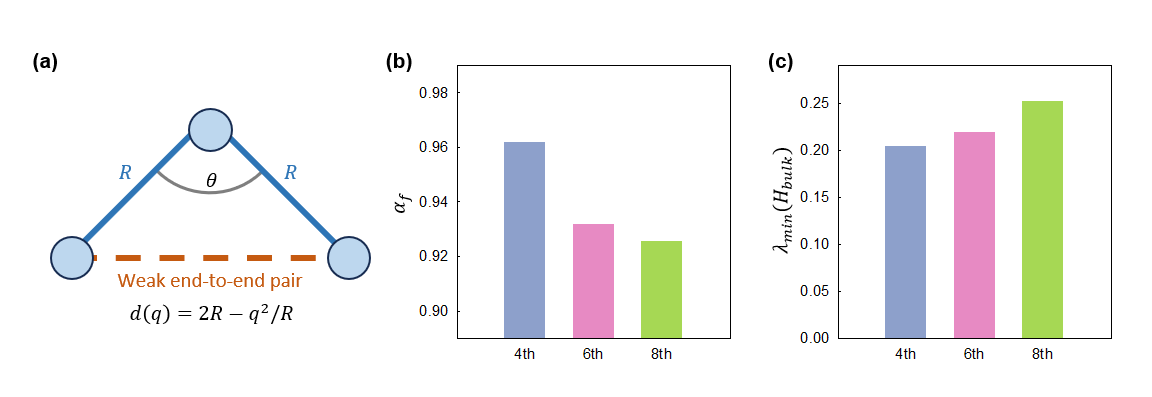}
\caption{Microscopic mechanism and finite-order robustness. (a) Schematic strong-bond two-arm geometry with a weak end-to-end pair interaction; not a molecular simulation or a proof of the complete coupled linkage. (b) Independently calculated interface folds for quartic, sixth-, and eighth-order structural potentials. (c) Positive smallest excited-bulk Hessian eigenvalues at the respective folds. These are numerical examples, not proof for arbitrary polynomial order.}\label{fig:robust}
\end{figure*}

\textit{Constrained mechanical realization and limits of generality.---}
The potential in Eq.~\eqref{eq:U} need not be interpreted as an inserted negative-stiffness spring. Two positive harmonic penalties sharing an accommodation coordinate $S$ yield
\begin{equation}
V(S,q)=\frac{k_1}{2}(S-a_1q)^2+\frac{k_2}{2}(S-b_1q^2)^2+\frac{\mu_0}{2}q^2,
\end{equation}
where $k_1,k_2>0$. Exact minimization over $S$ gives
\begin{equation}
U_{\rm eff}(q)=\frac{k_1k_2}{2(k_1+k_2)}(a_1q-b_1q^2)^2+\frac{\mu_0}{2}q^2.\label{eq:micro}
\end{equation}
Choosing $k_1=k_2=1$, $a_1=4/3$, $b_1=1$, and $\mu_0=\alpha-8/9>0$ reproduces Eq.~\eqref{eq:U} exactly. An explicitly constrained two-arm hinge supplies an exact quadratic chord response, while directional bonding can favor the required geometric constraints (Fig.~\ref{fig:robust}). This is an exact construction \emph{within the specified constrained Hamiltonian}, not a parameter-free derivation from an unconstrained molecular liquid; the simultaneous kinematic constraints and the mapping from molecular configurations to $q$ remain additional physical hypotheses.

Independently calculated sixth- and eighth-order positive structural extensions also exhibit localized folds before excited-bulk instability (Fig.~\ref{fig:robust}). The three finite polynomial examples support robustness under selected perturbations, not a theorem for arbitrary $U$. The exact constrained reduction, the hinge geometry, and the finite-order numerical robustness evidence are detailed in Supplemental Material, Secs.~S17--S19, respectively; the relation to the earlier CVFP and symmetry-field formulations is specified in Sec.~S20. In particular, a conditional continuation argument for small smooth perturbations requires nondegeneracy and appropriate continuum Fredholm and spectral hypotheses, which have not been proved for the infinite-domain system.

\textit{Physical implication.---}
A material-specific temperature singularity requires a constitutive path $(\chi(T),\alpha(T))$ that crosses the interface fold while $\chi(T)<2$. For the illustrative upper-consolute parametrization $\chi(T)=2T_c/T$, the one-phase condition is $T>T_c$ and the present reference value $\chi=1.90$ corresponds to $T/T_c=2/1.90$. The interface threshold is reached only if the independently specified structural control satisfies $\alpha(T_f)=\alpha_f[\chi(T_f)]$ along that path. For a transverse crossing on the flowing side, Eq.~\eqref{eq:scaling} conditionally predicts $t_{\rm pass}\propto|T-T_f|^{-1/2}$ with a finite equilibrium bulk correlation length. No such calibrated material trajectory or dimensional lifetime is claimed (Supplemental Material, Secs.~S14--S15). Interfacial soft modes in wetting and surface transitions provide useful comparators~\cite{lipowsky1982,nakanishi1982,bonn2009}; stochastic activation can regularize a deterministic divergence in an actual fluctuating liquid~\cite{hanggi1990}. Direct tests should compare prepared interfaces with independently prepared excited homogeneous controls and track both core loss and concentration-contrast loss.

The equilibrium compositional correlation length remains finite, $\xi_{\rm eq}\propto\sqrt{\kappa/f_{\rm eq}''}$, since $f_{\rm eq}''\geq0.20$. This differentiates the mechanism from a diverging equilibrium bulk Ornstein--Zernike length. The prediction is falsified in a candidate system if homogeneous excited controls lose stability at the same threshold as the interface, if the localized mode is absent, or if core passage remains nonsingular across the proposed threshold. The result is a numerically supported, spatially localized saddle-node mechanism in a strictly miscible equilibrium mixture: a divergence of deterministic interfacial passage time \emph{before} excited-bulk instability, coupled to an evolving concentration profile. Its claim is deliberately narrower than universal structural-potential behavior or a fully microscopic prediction for a particular fluid.

\begin{acknowledgments}
We thank Prof. Hiroshi Watanabe for insightful discussions. H.S.L. acknowledges support from Dong-A University, the Technology Innovation Program (No.~20010315), MOTIE (RS-2023-00258521), and the NRF Basic Science Research Program (RS-2023-00248126).
\end{acknowledgments}

\end{document}